\documentclass[10pt,conference]{IEEEtran}

\usepackage{amsmath,amssymb}
\usepackage{color}
\usepackage{graphicx}
\include{epsf}
\usepackage{dsfont}
\usepackage{bbm}

\def\(({\left(}
\def\)){\right)}
\def\[[{\left[}
\def\]]{\right]}

\newcommand{\brb}[1]{\left[#1\right]}

\newcommand{\be}{\begin{equation}}
\newcommand{\ee}{\end{equation}}
\newcommand{\bea}{\begin{eqnarray}}
\newcommand{\eea}{\end{eqnarray}}

\begin{document}

\title{Non-adaptive pooling strategies for detection of rare faulty items}
\author{
\IEEEauthorblockN{Pan Zhang}
\IEEEauthorblockA{CNRS and ESPCI \\ UMR 7083 Gulliver\\ 10
  rue Vauquelin\\ Paris 75005, France.}
\and
\IEEEauthorblockN{Florent Krzakala}
\IEEEauthorblockA{CNRS  and ESPCI \\  UMR 7083 Gulliver\\ 10 rue Vauquelin\\Paris 75005, France.}
\and
\IEEEauthorblockN{Marc  M\'ezard}
\IEEEauthorblockA{ENS, Paris and \\Univ. Paris-Sud/CNRS, LPTMS \\  B\^{a}t.~100, 91405\\
Orsay, France.
}
\and
\IEEEauthorblockN{Lenka
  Zdeborov\'a}
\IEEEauthorblockA{Institut de Physique Th\'eorique\\ IPhT, CEA Saclay\\ and URA 2306,
CNRS\\ 91191 Gif-sur-Yvette, France.}
}

\maketitle
\begin{abstract}
  We study non-adaptive pooling strategies for detection of rare
  faulty items.  Given a binary sparse $N$-dimensional signal $
  x$, how to construct a sparse binary $M \times N$ pooling matrix~$F$
  such that the signal can be reconstructed from the smallest possible
  number $M$ of measurements $y = F x$? We show that a very
  low number of measurements is possible for random spatially coupled
  design of pools $F$. Our design might find application in genetic
  screening or compressed genotyping. We show that our results are
  robust with respect to the uncertainty in the matrix $F$ when some
  elements are mistaken.
\end{abstract}

\section{Introduction}
Group testing \cite{du1993combinatorial,GroupTesting}, also known as
pooling in molecular biology, is designed to reduce the number of
tests required to identify rare faulty items. In the most naive
setting each item is tested separately and the number of tests is
equal to the number of items. If, however, only a small fraction of
the items are faulty, then the number of tests can be decreased
significantly by creating ``pools'', i.e. by including more than one
item in one test and allowing each item to take part in several
different tests.  The main problem is how to design these pools such
that their number is the smallest possible while allowing for a
tractable, and robust to noise, reconstruction procedure.

In adaptive group testing the new pools are designed using the
results of previous pools.  However, many experimental situations
require a non-adaptive testing where all pools must be specified
without knowing the outcomes of other pools. Mainly two kinds of tests
are relevant for practical applications. In Boolean group testing,
each test outputs negative if it does not contain a faulty item and
positive if it contains at least one faulty item. In linear group
testing, each test outputs the number of faulty items. In practical
applications the experimental constraints often require that the size
of each pool is relatively small and one item does not belong to too
many pools.  In this paper we analyse the non-adaptive (single stage)
pooling with linear tests and limited size of each pool.

A number of recent works have discussed a close relation between
non-adaptive pooling with linear tests and compressed sensing
\cite{GilbertIwen09,ErlichShental2009,Shental2009,Shental2010,Erlich:10,Kainkaryam:10,Narayanaswamy:11,Mourad:12}.
Here, we follow this direction of works and build on recent advances
in compressed sensing that used spatially coupled design of
measurements and permitted to decrease the number of
measurements down to the information theoretical limit
\cite{KrzakalaPRX2012,DonohoJavanmard11,KrzakalaJSTAT2012}.

Our results are meant to find applications in various currently
relevant problems e.g.  genetic screening \cite{ErlichShental2009} or
compressed genotyping \cite{Erlich:10}.  In these applications, genes
of several individuals are mixed together and one measures how many of
the genes in a given pool are "faulty'', e.g. related to certain
genetic problem. This is usually done by introducing markers that
attach to these faulty genes. The main source of noise in such
experiments is that the marker does not attach, i.e. whereas one
thought that a given gene belonged to a given pool, in reality it did
not. We shall investigate the robustness of our results under this
type of noise.

\section{Setting and related works}
The non-adaptive pooling for detection of rare faulty items that we
investigate is defined as follows: Consider a sparse binary
$N$-dimensional vector $ x$. Its components (items) are denoted by $x_i$,
$i=1,\dots,N$. The vector is sparse: only $\rho N$ of the components
are $x_i=1$ (faulty) and the others are $x_i=0$ (correct), with $\rho\ll
1$. A pool $a$ is a subset of components $a \subset \{1,\dots,N\}$. We
denote $F_{ai}=1$ if component $i$ belongs to pool $a$ and $F_{ai}=0$
otherwise. The result of a pool/test \be y_a=\sum_{i\in a} x_i =
\sum_i F_{ai} x_i \ee is the number of faulty items in the pool. The
goal is to design a smallest possible number $M$ of pools such
that the vector $ x$ can be reconstructed in a tractable way from the
results of these pools $ y=(y_1,\dots,y_M)$. In practice the result of
a pool becomes often unreliable if the pool contains many items and
also if one item belongs to many pools, because the sample
corresponding to one item then needs to be split into many small
pieces. We are hence interested in the case where the size of every
pool $K_a$ is small (compared to $N$) and every item belongs to only a
small number $L_i$ of pools.

This problem is reminiscent of compressed sensing
\cite{CandesTao:05,Donoho:06} which is designed to measure signals
directly in their compressed form. In fact our problem is compressed
sensing with the additional constraint that the signal is binary and
with a sparse binary measurement matrix $F_{ai}$. We also consider
matrix uncertainty: some elements assumed to be $F_{ai}\!=\!1$ are in
fact $F_{ai}\!=\!0$ with probability $p$. The field of low-density
parity check (LDPC) error correcting codes \cite{Gallager62} provides
information about sparse measurement matrices with which tractable
reconstruction can be achieved. Indeed, the only difference between
non-adaptive group testing with linear tests and LPDC is that the
algebra is over integers in group testing instead of $GF(2)$ in
LDPC. The spatially coupled pooling design we study here was first
discovered and validated in the field of error correcting codes
\cite{FelstromZigangirov99,LentmaierSridharan10,LentmaierMitchell10,KudekarRichardson10}.

The reconstruction algorithm that is most commonly used in compressed
sensing and that has been also discussed several times for
reconstruction in group testing and pooling experiments
\cite{GilbertIwen09,Shental2009,Shental2010} is based on a linear
relaxation of the problem to real signal components $0 \le x_i \le
1$. One then minimizes the $\ell_1$-norm of the signal under the
constraints $ y = F x$. This is a convex problem that can be solved
efficiently using linear programing. In what follows we will use the
$\ell_1$ reconstruction as a reference benchmark to demonstrate the
improvement that can be achieved using our pooling design and the
belief propagation based reconstruction.

The problem of non-adaptive group testing with linear tests, but with
no constraints on the sparsity of the matrix $F$, is also known as the
{\it coin weighting problem} \cite{Bshouty09,MarcoKowalski12}.  A
detailed review of previous results can be found in
\cite{Bshouty09}. It was shown \cite{ErdosRenyi63} that, in the limit
that interests us where $N\to \infty$, reconstruction is not possible
with less than $M = (2\log{2}) N/\log{N}$ tests. On the other hand a
successful deterministic construction of the measurement matrix,
together with a polynomial reconstruction algorithm, was found with $M
= (2\log{2}) N[1+o(1)]/\log{N}$ \cite{Lindstrom64} measurements. As far
as we know, however, the problem with a small (constant) number of
items in every pool, as treated here, is still open.

Note that when the number of faulty items $R$ is much smaller than
$N$, then another line of works should be considered. The best
polynomial time non-adaptive algorithms known for the coin weighting
problem then need $M=R \log_2{N}$ measurements
\cite{Lindstrom72}. Our approach is thus useful only in regimes where
the number of faulty items is larger than $R>
2(\log{2})^2N/\log^2{N}$. Another problem that is closely related to
non-adaptive pooling as considered here is the sparse code division
multiple access (CDMA) method \cite{RaymondSaad07,GuoWang08}; with the
difference that the signal $x$ is {\it not} sparse and that there is
usually a considerable Gaussian additive noise on the measurement
vector $y$. The goal in CDMA is not to minimize the number of
measurements $M$ (the number of chips), but to support the largest
possible amount of noise.  Spatial coupling was investigated for
(dense) CDMA in \cite{Schlegel2011,Takeuchi2011}.

\section{Spatially coupled design of pools}
\begin{figure}[!ht]
 \centering
\includegraphics[width=0.24\textwidth]{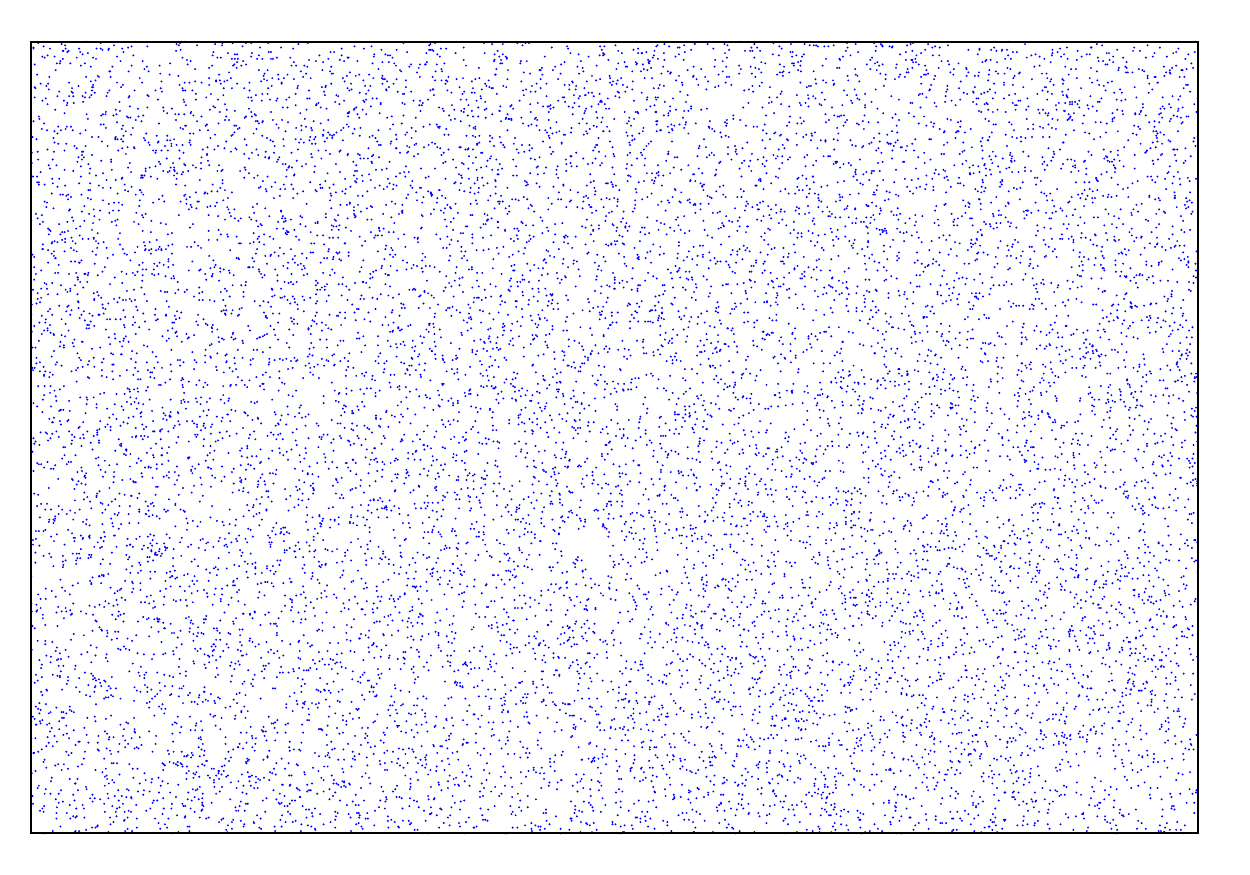}
\includegraphics[width=0.24\textwidth]{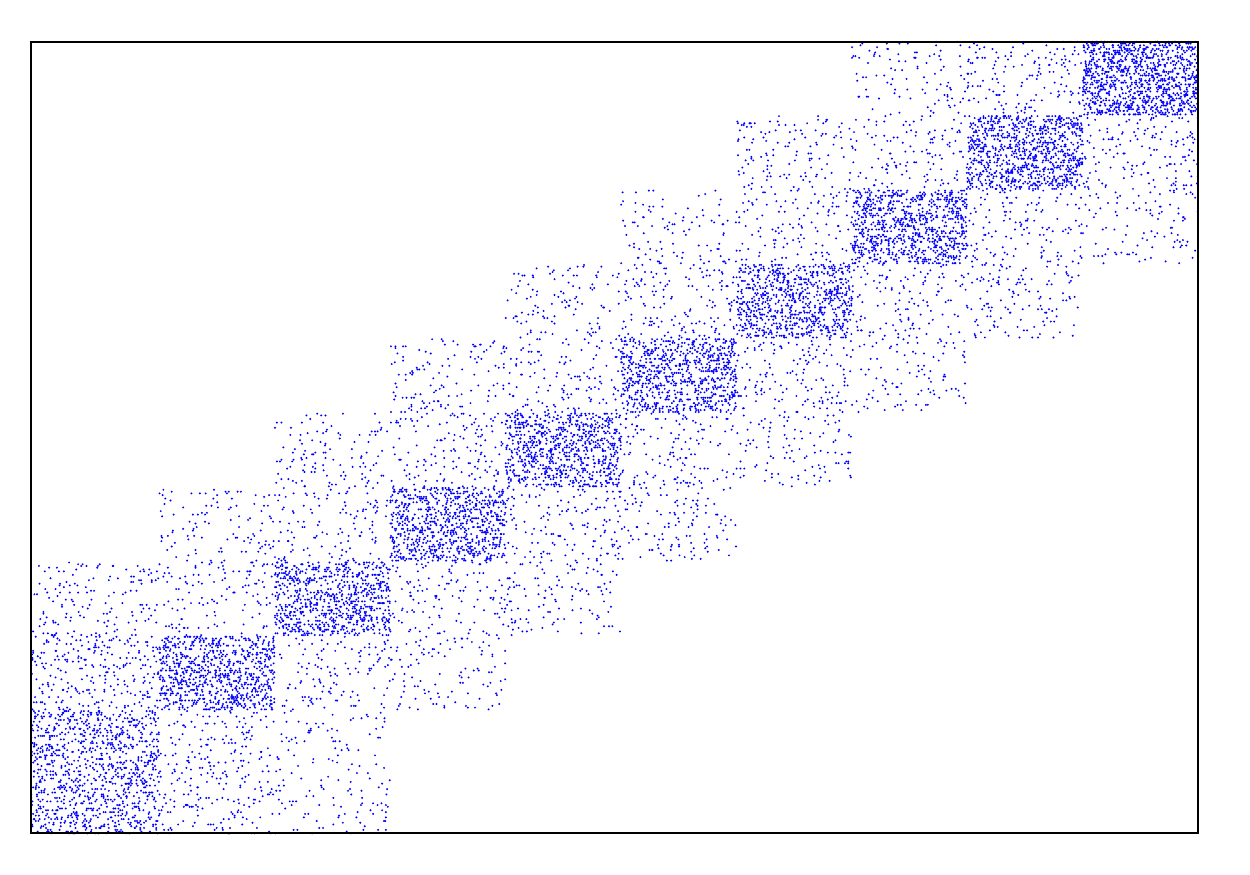}
\caption{\label{fig:matrices} Left: Random pooling design, each blue
  point corresponds to an item $i$ belonging to a pool $a$. We took
  $N\!=\!2000$ items, $M\!=\!700$ pools, each item participates in $L\!=\!7$
  pools, and each pool has $K\!=\!20$ items.  Right: Seeded (or spatially
  coupled) design of pools, with $N\!=\!2000$ items, $M\!=\!490$ pools,
  $B\!=\!10$ blocks with $K_s\!=\!20$, $K_f\!=\!30$, $J\!=\!0.2$ and $w\!=\!2$.}
\end{figure}

The first design of pools (${\cal D}_r$) that we consider is based on
a random construction. Each pool $a$ contains $K$ items. Each item $i$
belongs to $L$ pools. The assignment of items into pools is chosen
uniformly out of all such possible ones. An example of a such random
pooling matrix is plotted in Fig.~\ref{fig:matrices} left.

The second design of pools (${\cal D}_s$), for which the total number
of tests needed for successful reconstruction will be considerably
smaller, is the ``seeded'' or ``spatially-coupled'' design,
illustrated in Fig.~\ref{fig:matrices} right.  First, we divide the
$N$ items into $B$ equally sized blocks. Pools are also divided into
$B$ blocks, the first of them (the ``seed'') being larger than the
others. We will fix to $L$ the degree of all items, to $K_s$ the
degree of pools in the first block, and to $K_f$ the degrees of pools
in the other $B-1$ blocks. The size of the first block is then
$M_s\!=\!NL/(K_sB)$, and that of the other blocks
$M_f\!=\!NL/(K_fB)$. The overall under-sampling ratio is then
$\alpha\!=\!L/(K_sB) + (B-1)L/(K_fB)$.  When deciding connections
between items and pools, we first connect randomly each pool to items
in the block with the same index.  Then we apply the following
rewiring procedure: with probability $J$ for each edge
(i.e. connection between an item and a pool), we chose randomly
another edge whose item is in one of the $w$ previous blocks (and that
has not been rewired yet) and switch the two edges. The details of
this rewiring procedure do not change our results as long as a
fraction of about $J$ of new connections is created up to distance
$w$.

\section{Upper and lower bounds for number of pools}

A simple lower bound on the number of necessary measurements $M$ can
be obtained as follows: To reconstruct exactly the $N$-component
signal of $R=\rho N$ non-zero components each possible configuration
of the signal $ x$ should correspond to a distinct result of the pools
$ y$.  In other words the number of possible outcomes of the
measurements must be larger than the number of possible signals. If
$K$ denotes the number of items in each test this gives ${N \choose R}
\le (K+1)^M$. In the limit of large systems, and constant density of
faulty items $\rho$ this gives $ \alpha_{\rm LB} = H(\rho) / \log{(K+1)}
$, where we use the entropy function $H(\rho) =-\rho
\log{\rho}-(1-\rho)\log(1-\rho)$.  Note that this lower bound holds
also for adaptive pooling design.

We also derived a first moment upper bound on the critical ratio above which
reconstruction is in principle possible for the random design (${\cal D}_r$). Calling $\cal N(\epsilon)$ the number of vectors $x$ compatible with the
measurements $y$ at distance $1-\epsilon$ from the original signal, one
obtains:
\bea
\mathbb{E}({\cal N(\epsilon)})  \!\!\!&=&\!  e^{N\Phi(\rho,\epsilon)}  = e^{N[\alpha \log{P(\epsilon,K,\rho) } +H(\epsilon,\rho) 
  ]}, \!\label{eq:fm:2} \\ 
\text{where  }
  H(\epsilon,\rho) 
\!\!\!  &=& \!  -\epsilon \rho \log{\epsilon \rho}- 2 (1-\epsilon) \rho
   \log{(1-\epsilon) \rho} \nonumber\\ && \!\!\!-  (1-2\rho + \epsilon \rho) \log{(1-2\rho
     + \epsilon \rho)}- H(\rho), \nonumber \\
\text{and }
   P(\epsilon,K,\rho) \!\!\!&=& \!\!\! \sum_{a=0}^{  \frac{K}{2} }    \frac{K!  \left[  1-2\rho(1-\epsilon )
   \right]^{K-2a}   [(1-\epsilon) \rho]^{2a} }{(a!)^2
   (K-2a)!} \, . \nonumber
\eea
When $\Phi(\rho,\epsilon)$ is negative everywhere except in the
vicinity of $\epsilon\!=\!1$ then the linear system $ y = F x$ has
only one solution that corresponds to the signal to be
reconstructed. This leads to a threshold value $\alpha_{\rm UB}$ above
which reconstruction is in principle possible. As we will see this
upper bound is very close to the actual threshold.  With $K\to\infty$,
one can show that $\alpha_{\rm UB}=2\frac{H(\rho)}{\log K}$. For dense
matrices $F$ with $K=N$, this is a tight bound, as mentioned in the
context of coin weighting.

\section{Signal recovery algorithms}
Knowing the set of tests or pools, i.e. the binary matrix $F$, and the
measurement results $y$, one wants to reconstruct the signal. In
compressed sensing the algorithms that achieve exact reconstruction
with a smallest possible number of measurement are based on Bayesian
inference, see \cite{KrzakalaPRX2012,DonohoJavanmard11}. We thus adopt
the same strategy here.

The probability distribution of observing measurements
$  y$ given the signal $  x$ and set of pools $F$ is given as 
\be
     P(y|x,F) = \prod_{a=1}^M   {\sum_{i\in a} x_i  \choose y_a}  (1-p)^{y_a} p^{(\sum_{i\in a} x_i -y_a)}
\ee
To estimate $P(x|y,F)$, with the use of Bayes rule $P(x|y,F) =
P(y|x,F) P(x|F)/ P(y|F) $, we need to assume some knowledge of
statistical properties of the signal $x$. Denoting the fraction of
non-zero elements $\rho$ we use a prior 
\be P(x) =
\prod_{i=1}^N\brb{(1-\rho)\delta(x_i)+\rho \delta(x_i-1)} \ee
In compressed sensing it was argued \cite{DonohoJavanmard11} that for
pooling matrices with random elements this assumption works as well
for iid signal components as for correlated signals. Of course if an
additional knowledge about the correlations in the signal were
available it could be exploited and the performance improved further.
We shall, however, not assume any such additional knowledge.  The
value of $x_i$ that minimizes the number of errors is obtained as the
value that is more probable according to the marginal distribution for
that element 
\be
\mu (x_i)=\sum_{ \{x_j \}_{j\neq i} } P(x|y,F).
\ee

To estimate these marginals we use the canonical
belief propagation (BP) algorithm \cite{YedidiaFreeman03}. Following
the usual derivation we introduce for each non-zero matrix element
$F_{ai}$ two messages, $\chi^{i\to a} $ and $\psi^{a \to i}$,
which are two-component vectors (normalized to $\chi^{i\to
  a}_0+\chi^{i\to a}_1=1$ and $\psi^{a \to i}_0+\psi^{a \to i}_1=1 $), and we write
iterative update equations for these as
\bea 
\chi^{i\to a}_{x_i}&=&\frac{\rho^{x_i}(1-\rho)^{1-x_i}\prod_{b\in\partial i\backslash
    a} \psi^{b\to i}_{x_i} }{\rho \prod_{b\in\partial i\backslash a}
  \psi^{b\to i}_{1} + (1-\rho)\prod_{b\in\partial i\backslash a} \psi^{b\to
    i}_{0}} \, ,\label{BP_bin1}\\
\psi^{a \to i}_{x_i}&=& \frac{1}{Z^{a \to i}}
\sum_{B=y_a-x_i}^{K-x_i} {B+x_i \choose y_a} (1-p)^{y_a} p^{B+x_i-y_a}\nonumber\\
&& \sum_{{\{ {x_j}\}_{j\in \partial a\backslash i}, \sum_{j}  x_j = B}}
 \, \, \prod_{j\in \partial a\backslash i} \chi^{j\to a}_{x_j}\, . \label{BP_bin2}
\eea
In the iterative BP algorithm we initialize $\psi_1$ and $\chi_1$ as
random numbers from the interval $(0,1)$ and update equations
(\ref{BP_bin1}-\ref{BP_bin2}) till convergence. Note that the
argument in equation
(\ref{BP_bin2}) depends only on the sum of variables, it can thus be
updated  with the use of a convolution in $(K_a-1)^2/2$ steps (compared to the naive $2^{K_a-1}$ steps).  
Once convergence is
reached the BP estimates of the marginal probabilities are computed
by:
\be 
\chi^{i}_{x_i}=\frac{\rho^{x_i}(1-\rho)^{1-x_i}\prod_{a\in\partial i} \psi^{a\to i}_{x_i} }{\rho \prod_{b\in\partial i\backslash a}
  \psi^{a\to i}_{1} +(1-\rho) \prod_{a\in\partial i} \psi^{a\to i}_{0}}\, .
\ee
The BP inference of the item $i$ is then $x_i^* =1$ is $\chi^i_1 >
\chi^i_0$ and $x_i^* =0$ otherwise. 

The iteration of message updates, starting from random messages, is
the BP algorithm.  It turns out that BP can also be used to analyse
the optimal Bayes inference. For the purpose of this analysis one
initializes BP messages to values corresponding to the true signal,
and iterates the equations until convergence. In some region of
parameters this will reach a different fixed point than the iterations
of randomly initialized messages. The fixed point that corresponds to
the result that would be achieved by the (exponentially costly)
optimal Bayesian inference procedure is the one having the largest
log-likelihood $\Phi$. The BP estimate of the log-likelihood, also
called Bethe free energy, is given by \bea
\Phi&=&\sum_i \log Z_i-\sum_a(K_a-1)\log Z_a\, , \label{eq:f}\\
Z_i &=& \rho \prod_{b\in\partial i\backslash a} \psi^{a\to i}_{1}
+(1-\rho) \prod_{a\in\partial i} \psi^{a\to
  i}_{0}\, , \nonumber \\
Z_a&=&\sum_{B=y_a}^{K} {B \choose y_a}
\left(\frac{1}{p}-1\right)^{y_a} p^{B} \sum_{\substack{ \{
    {x_j}\}_{j\in \partial a} \\ \sum_{j} x_j = B }}
\prod_{j\in \partial a} \chi^{j\to a}_{x_j} \nonumber \eea

In most practical situations the fraction of "faulty" items $\rho$ is
not known in advance. In such cases it can be learnt via expectation
maximization learning, by iterating the following expression
$\rho_{\rm new} = \sum_{i=1}^N \chi^i_1 /N$,
where the r.h.s. is evaluated using the previously estimated value of $\rho$.

\section{Performance and phase diagrams}

\subsection{Noiseless case}
\begin{figure}[!ht]
\centering
\includegraphics[width=0.4\textwidth]{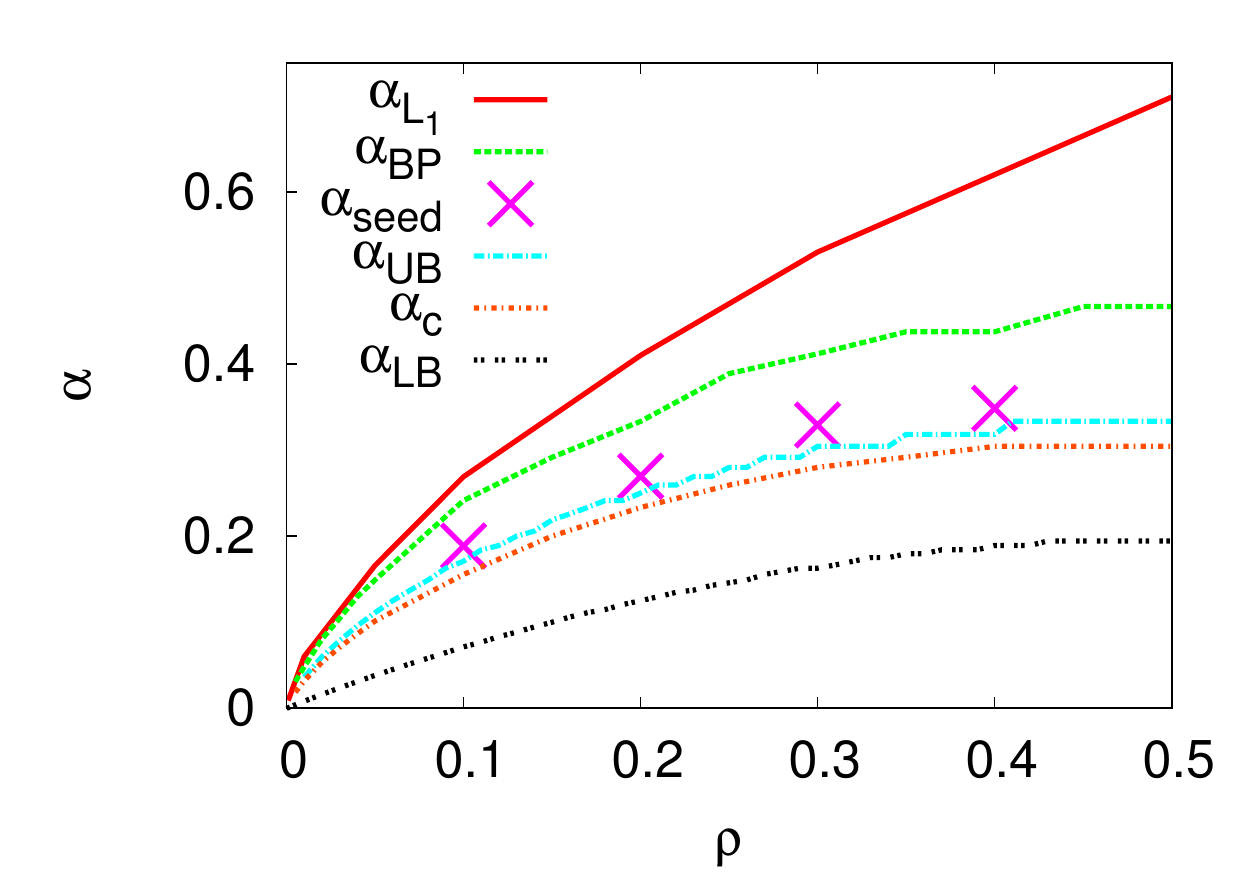}
\caption{\label{fig:compare} 
Lines are for random pooling design. From top: performance limit $\alpha_{\ell_1}$ for $\ell_1$,
performance limit $\alpha_{BP}$ for BP, first moment upper bound on
the Bayes-optimal inference, Bayes-optimal inference transition, and
information theoretic lower bound. Number of items is $N=10^5$, each
item is in $L=7$ pools. Data
points mark the best performance we achieved with seeded matrices with
$B=20$ and $w=2$, other parameters are listed in Table~\ref{tab:parameters}.
}
\end{figure}

Let us first investigate reconstruction of the signal for the random
design ${\cal D}_r$. We are interested in the smallest possible
ratio $\alpha_c=M_c/N$ for which exact reconstruction
is still possible in the large $N$ limit. With the BP algorithm, exact reconstruction is possible
if and only if $\alpha>\alpha_{BP}$.  The threshold (or
``phase transition'') $\alpha_{BP}$ is plotted in
Fig.~\ref{fig:compare}. We compare this value to the smallest
possible ratio $\alpha_{\ell_1}$ for which the standard convex
optimization approach, where one minimizes $|x|_{\ell_1}$ subject to
$  y = F   x$ and $0\leq x_i\leq 1$, provides exact
reconstruction.  We see in Fig.~\ref{fig:compare} that
$\alpha_{\ell_1}$ is only slightly
larger than $\alpha_{BP}$ for all range of $L$. 

\begin{figure}[!ht]
 \centering
\hspace{-0.4cm}
\includegraphics[width=0.26\textwidth]{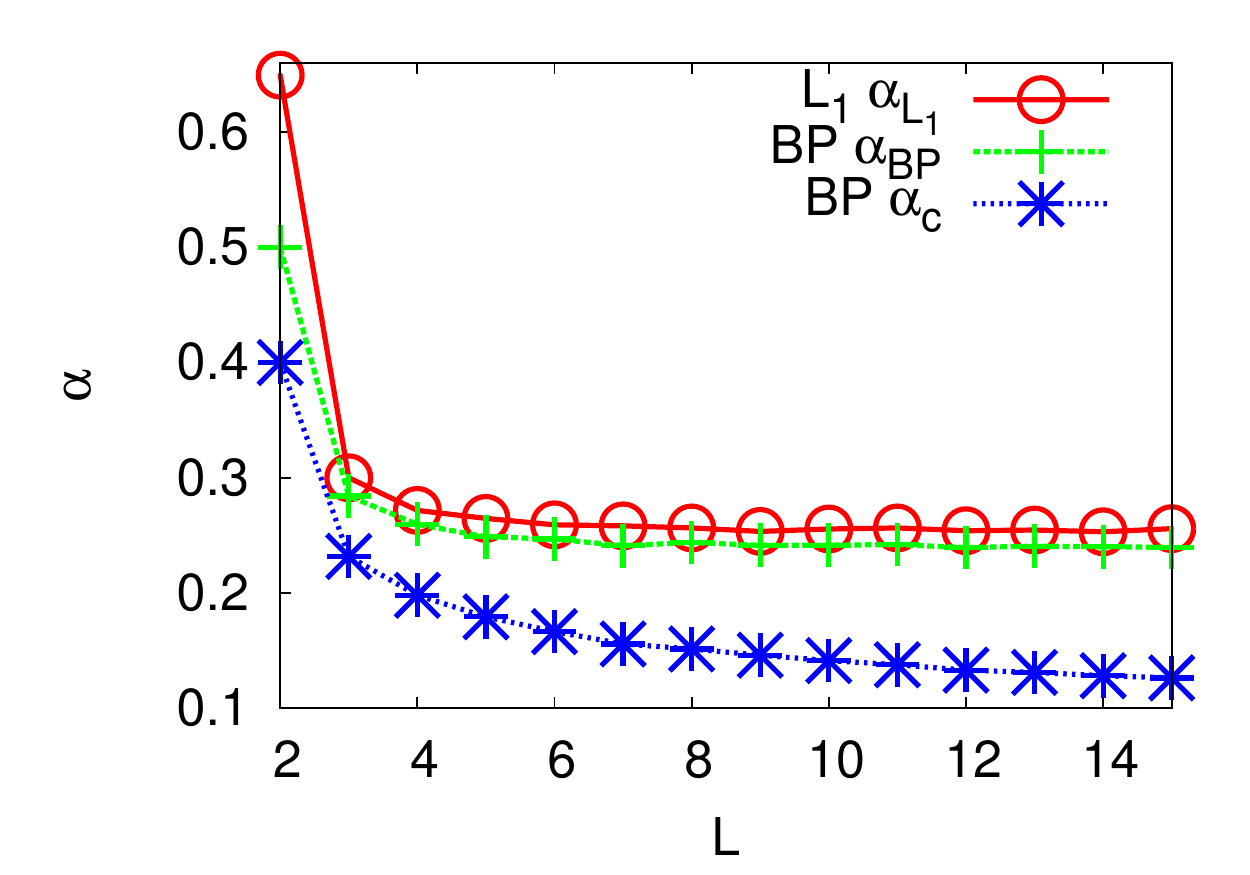}
\hspace{-0.6cm}
\includegraphics[width=0.26\textwidth]{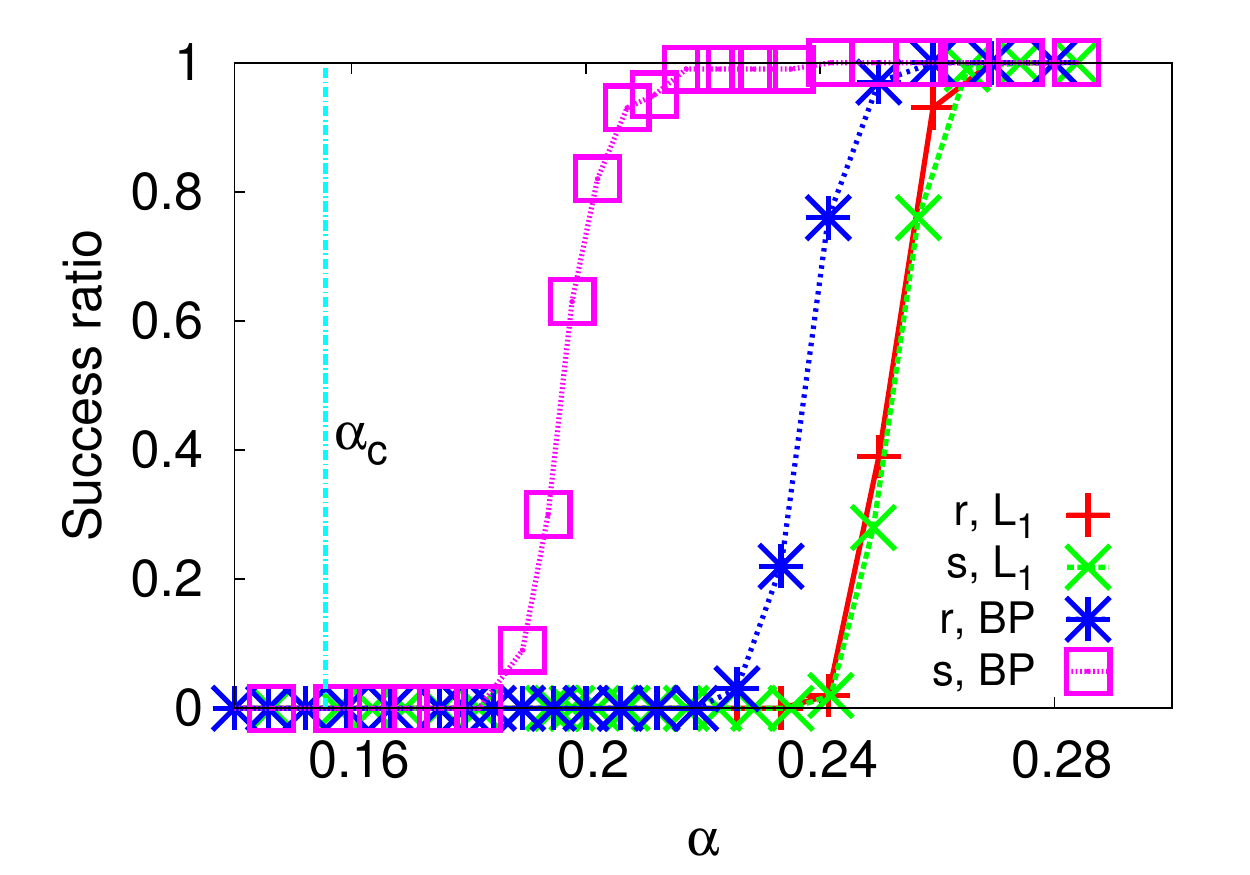}
\caption{\label{fig:randompool:alpha}
Left: The limit of performance for BP ($\alpha_{BP}$), $\ell_1$
minimization ($\alpha_{\ell_1}$) and the Bayes optimal inference
($\alpha_c$) for random design of pools. The values are obtained as
averages over $20$ instances with $N=10^4$ items and density of faulty
items $\rho=0.1$. Right: Fraction of exactly reconstructed signals
using BP and $\ell_1$ reconstruction with random and seeded pools for
$N=10^4$, $L=7$ and $\rho=0.1$. Data obtained from $100$ random
instances. The seeded matrix has $B=20$ blocks, the first block (seed)
has $K_s=20$, and following blocks have $K_f>20$ corresponding to
different total values of $\alpha$.  Fraction $J=0.4$ of links are
connected to {\bf $w=2$} previous blocks.
}
\end{figure}

Using the method explained  in previous section, we have investigated the performance of Bayes optimal
approach by evaluating the Bethe free energy $\Phi$. BP messages
initialized on the true signal are fixed point in the absence of
noise, corresponding to $\Phi=0$. Randomly initialized BP reaches the
same fixed point at large values of $\alpha$, at
$\alpha_{BP}$, $\Phi$ jumps discontinuously to negative values,
meaning that there is no other signal with density $\rho$ satisfying
all the tests. The log-likelihood then grows as $\alpha$ decreases and
becomes positive at $\alpha_c$ below which there are other signals of
density $\rho$ satisfying the tests and hence the true signal is
undetectable. Above $\alpha_c$ the exactly evaluated Bayes-optimal
inference would hence be able to reconstruct exactly the signal. 
The value of $\alpha_c$ is also plotted in
Fig.~\ref{fig:compare} and we see that both $\ell_1$ and BP for random
design of pools ${\cal D}_r$ are
considerably suboptimal.

Using seeded pooling design improves BP
performance by moving the ratio $\alpha$ above which BP is able to
reconstruct exactly the signal down to the Bayes-optimal threshold
$\alpha_c$. Several works have argued that quite generically when the system size $N$, the
number of block $B$ and the interaction range $w$ go to infinity (in
this order) BP with spatially coupled design is able to saturate the
threshold $\alpha_c$ \cite{LentmaierSridharan10,LentmaierMitchell10,KudekarRichardson10,KrzakalaPRX2012,DonohoJavanmard11}. This statement applies also to our
case. Here we investigate the performance of BP for seeded pooling
design with realistic values of parameters $N$, $B$, $w$. In
Fig.~\ref{fig:randompool:alpha} right we plot the fraction of
instances in which the signal was reconstructed exactly by $BP$ and
$\ell_1$ for the random pooling design ${\cal D}_r$, and for the
seeded design ${\cal D}_s$, with a
set of parameters specified in Table~\ref{tab:parameters}. 
Whereas
$\ell_1$ performs basically the same for both designs, the performance of BP
improves considerably in the seeded pooling scheme ${\cal D}_s$, for
realistic values of the parameters. 
In Fig.~\ref{fig:compare} we summarize all our results for the
noiseless case with 
$L=7$, and with a varying fraction of faulty items $\rho$. Let us also note
that the Bayes-optimal transition for seeded matrices
is not exactly equal to the one for random matrices, the two are,
however, so close that the difference is not distinguishable in
Fig.~\ref{fig:compare}.

\begin{figure}[!ht]
  \centering
\hspace{-0.4cm}
\includegraphics[width=0.26\textwidth]{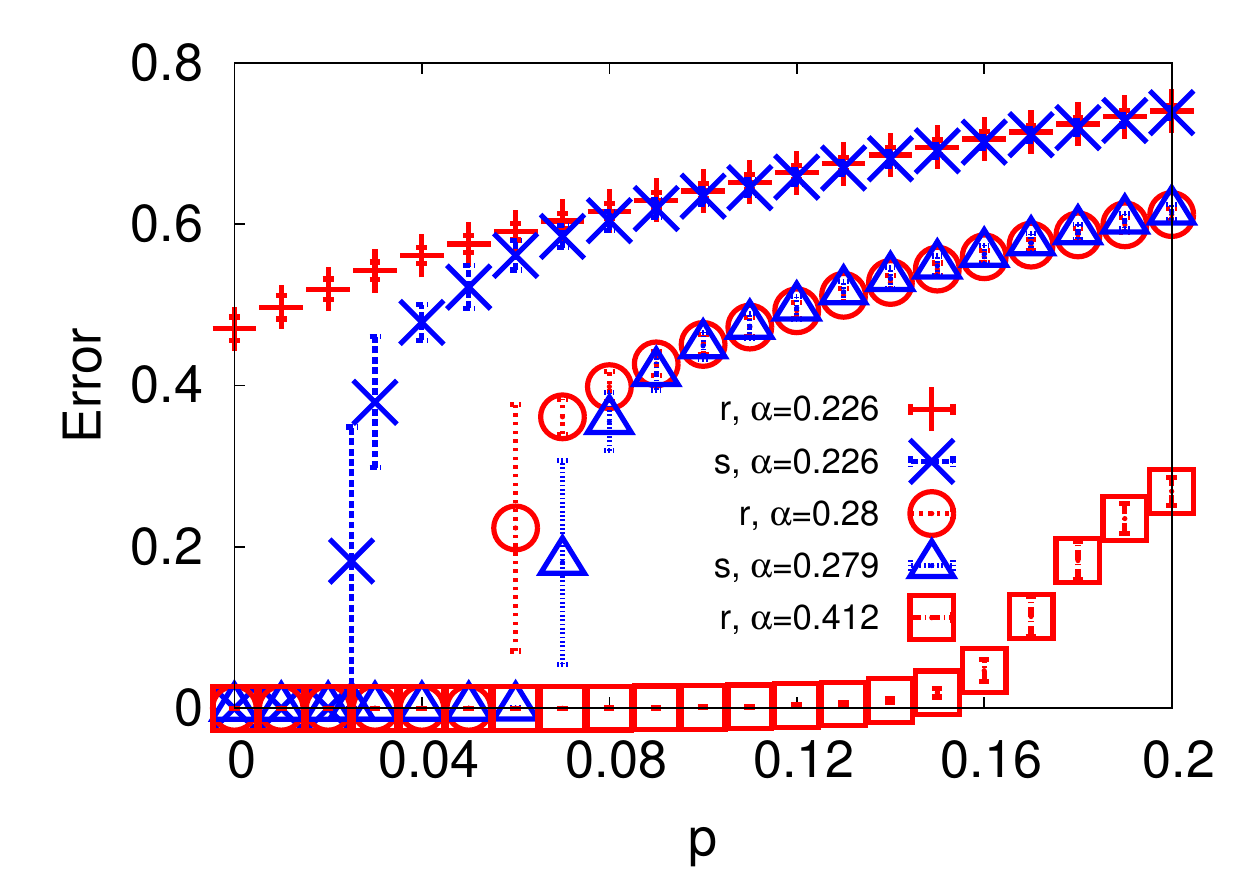}
\hspace{-0.6cm}
\includegraphics[width=0.26\textwidth]{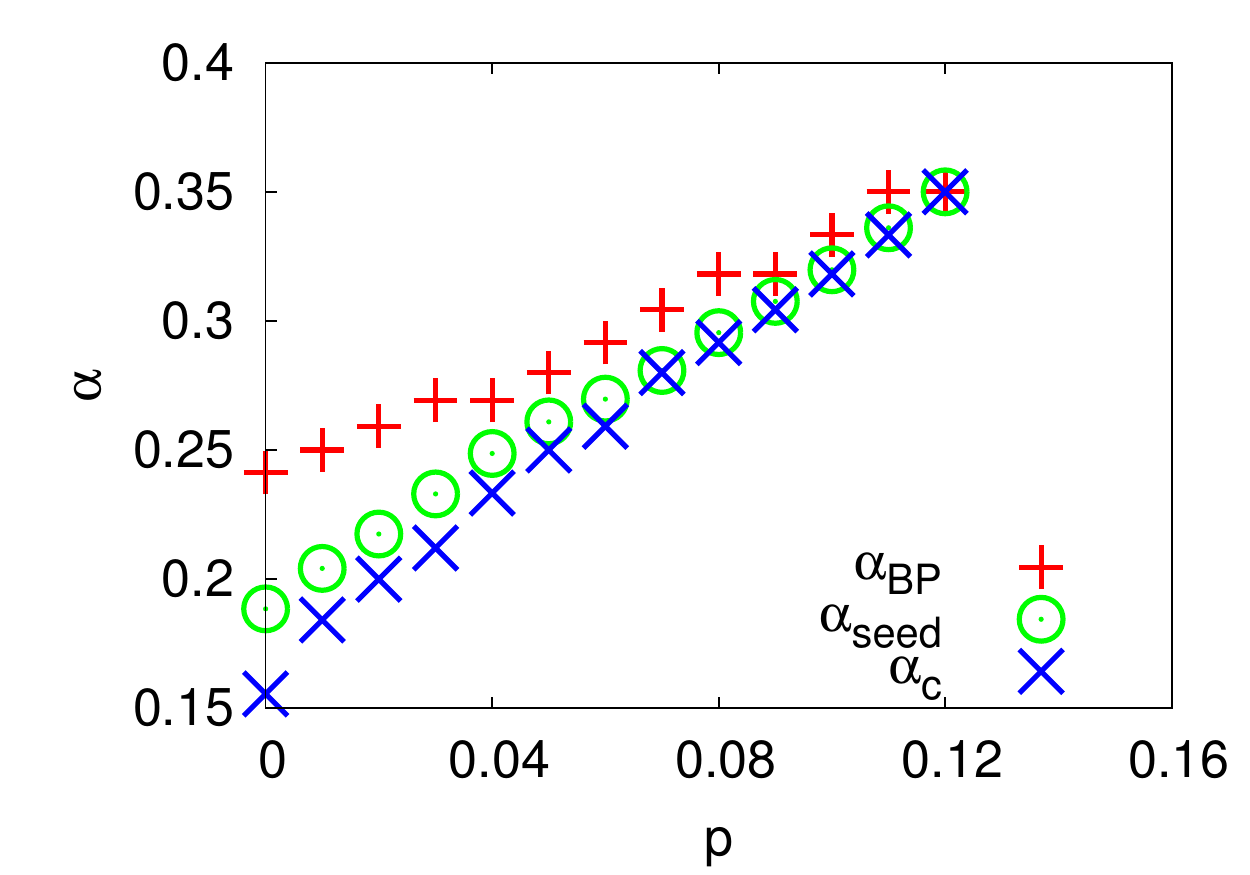}
\caption{\label{fig:noise:2}
Left: Error (fraction of wrongly-reconstructed items) as function of
noise strength $p$ for BP with random matrices and seeded matrices
with $N=5\times 10^4$, $\rho=0.1$, $L=7$, and different $\alpha$ values, each point is averaged over $20$ instances. Seeded matrices have $B=10$ blocks, parameters are listed in Table \ref{tab:parameters}.
Right: Spinodal, static transition and critical values for seeded matrices with $N=5\times 10^4, \rho=0.1, L=7$.
Seeded matrices have $B=20$ blocks, and parameters are listed in Table \ref{tab:parameters}.
}
\end{figure}

\subsection{Noisy case }

We investigate now the robustness of our results to measurement-matrix
noise. We denote by $p$ the probability that a matrix
element $F_{ai}=1$ was in fact $F^{\rm true}_{ai}=0$, i.e. item $i$
did not contribute to the result of the test $a$. 
With nonzero values of $p$ and large system size $N\to \infty$ exact
reconstruction is never possible, there is always a nonzero
probability $p^L$ that a faulty item was never included in any
measurement. However, for realistic sizes and small values of $p$ we
may still obtain exact or close-to-exact reconstruction. 

In the left part of Fig.~\ref{fig:noise:2} we plot the average error (fraction of wrongly
reconstructed items) as a function of the noise strength $p$. We see
that there is a critical value of $p$ above which the performance
deteriorates significantly. This value is larger for the seeded pooling design
than for the random pooling design. In the right part of
Fig.~\ref{fig:noise:2}  we then plot the critical values of ratio $\alpha$ as a
function of noise strength $p$ for
the random pooling design $\alpha_{BP}$, for the best seeded pooling
design $\alpha_{\rm seed}$ that we
found with realistic parameters, and for the Bayes optimal
reconstruction $\alpha_c$.  For $\alpha>0.35$
(for $\rho=0.1$, $L=7$) there
is no longer a value of $p$ where the performance deteriorates
sharply, instead the transition is smooth. Such a phase diagram is
qualitatively similar to the one of compressed sensing with
other types of noises, see e.g.~\cite{KrzakalaJSTAT2012,BarbierKrzakala12,KrzakalaMezard13}.

\begin{table}[!ht]
 \centering
	\caption{\label{tab:parameters}}
      \begin{tabular}{|l|l|l|l|l|l|l|l|l|}
		        \hline
				\multicolumn{9}{ |c| }{Parameters of seeded matrices in Fig.~\ref{fig:compare}} \\
		        \hline
				        $\rho$  & $K_s$ & $K_f$ & $J$ & &$\rho$  & $K_s$ & $K_f$ & $J$ \\ 
						\cline{1-4}
						\cline{6-9}
						        0.1    &  20    & 39     & 0.1 & & 0.2    &  15    & 27     & 0.1        \\ 
						        0.3    &  13    & 22     & 0.1 & &0.4    &  11    & 21     & 0.1      \\ 
		        \hline
		        \hline
				\multicolumn{9}{ |c| }{Parameters of seeded matrices in Fig.~\ref{fig:noise:2} left} \\
		        \hline
				        $\alpha$  & $K_s$ & $K_f$ & $J$ & &$\alpha$  & $K_s$ & $K_f$ & $J$ \\ 
						\cline{1-4}
						\cline{6-9}
						        0.226    &  20    & 33     & 0.1 & & 0.279    &  19    & 26     & 0.1        \\ 
		        \hline
		        \hline

				\multicolumn{9}{ |c| }{Parameters of seeded matrices in Fig.~\ref{fig:noise:2} right} \\
		        \hline
				        $p$  & $K_s$ & $K_f$ & $J$ & &$p$  & $K_s$ & $K_f$ & $J$ \\ 
						\cline{1-4}
						\cline{6-9}
						        0    &  20    & 39     & 0.1 & & 0.01    &  18    & 36     & 0.2        \\ 
						        0.02    &  16    & 34     & 0.3 & &0.03    &  19    & 31     & 0.3      \\ 
						        0.04    &  18    & 29     & 0.3 & &0.05    &  24    & 27     & 0.3       \\ 
						        0.06    &  15    & 27     & 0.4 & &0.07    &  14    & 26     & 0.4     \\ 
						        0.08    &  19    & 24     & 0.3 & &0.09    &  19    & 23     & 0.4    \\ 
						        0.1    &  20    & 22     & 0.4  & &0.11    &  20    & 21     & 0.4   \\ 
						        0.12    &  20    & 20     & 0.4  & & & & &\\ 
\hline
	\end{tabular}
  \end{table}

\section{Conclusion}
We have studied non-adaptive pooling strategies for detection of
rare faulty items. We have shown that the belief-propagation
reconstruction algorithm, together with a seeded (spatially-coupled)
design of the pools, leads to the best-known performance so far in the
sense that it minimizes the number of measurements necessary for exact
reconstruction in the noiseless case. Our results are very close to
Bayes optimality and robust with respect to measurement noise
corresponding to a faulty knowledge of the pools.

It is quite possible that this pooling design and its
reconstruction algorithm will find applications in genetic
screening. We note that our work can be extended to the case when the
non-zeros items in the signal are real-valued. In this case the BP
algorithm needs to be replaced by an AMP type of algorithm
\cite{DonohoMaleki09}. We are currently investigating this case for
sparse measurement matrices.

\section*{Acknowledgment}
This work has been supported in part by the ERC under the European
Union’s 7th Framework Programme Grant Agreement 307087-SPARCS, by the
EC Grant ‘‘STAMINA’’, No. 265496, and by the Grant DySpaN of
‘‘Triangle de la Physique.’’

\bibliographystyle{IEEEtran}
\bibliography{refs}
\end{document}